# Porosity properties of porous ceramic substrates added with zinc and magnesium material

D. Bouras[a], A. Mecif[a], R. Barillé[b], A. Harabi[c] and M. Zaabat[a]

[a] Laboratory of Active Components and Materials, Larbi Ben M'Hidi University, Oum El Bouaghi 04000, Algeria.

[b]MOLTECH-Anjou, Université d'Angers/UMR CNRS 6200, 2 Bd Lavoisier, 49045 Angers, France.

[c]Ceramics Laboratory, Mentouri University, Constantine 25000, Algeria.

## Abstract

Ceramic based materials covered with a mixture of mullite and zircon on the top, strengthened by the addition of zinc and magnesium compounds at different rates (37 % and 50 %) and made by a co-precipitation method are studied. The thermally treated and shaped materials prepared in the shape of cylindrical samples were modified to obtain different porosities. The addition of zinc to the ceramic based material leads to a significant increase and control of the density of porosity compared to the addition of magnesium. The porosity of all the materials was characterized. The permeability characteristics and the ability for all the pellets to absorb water until the saturation were also studied as a function of the porosity.

## 1.Introduction

The development of new microporous ceramic membranes [1 - 4] for nanofiltration [5 - 6], ultrafiltration [7 - 8] and microfiltration [9 - 10] have been studied since several years in the goal to obtain cheaper finished products. The use of organic membranes is currently more developed, but in some specified applications ceramic membranes have a number of advantages, such as a better mechanical, thermal and chemical resistance [11].

Porous ceramics have found many interests and can be used as membranes, filters, catalyst supports, heat exchangers, supports for bone regeneration, sensors, thermal insulators in construction or in aerospace applications, flue gas burners and lightweight materials for construction [12-14]. Membranes on the other hand are used in clarification and concentration operations in various fields such as: agribusiness, biotechnology, pharmacy, water and effluent treatment. Membrane processes are often economically competitive compared to traditional processes and may also contribute to the improvement of other process-coupled separation techniques [15].

The originality of this present work lies in the use of an inexpensive and abundant material for the membrane fabrication. The basic component is obtained from a local Algerian region of the north-east of Algeria (Guelma). The material is composed of Kaolinite and halloysite and is rich of Alumine. The ceramic material is processed with mullite and zircon and mixed with or without the addition of zirconia, after an adequate heat treatment at 1200°C [16- 17].

Moreover, in order to increase the material absorption or permeability by controlling pore diameters, other compounds were added to the ceramic based material. Magnesium and Zinc were chosen for mixing materials. We have found that in most of the time, the absorption of the overlapping material is larger than the unsupported ceramic material, especially at high temperatures [18 - 19]. In this work, the pellets are made with prepared modified ceramics and tested with distillate water ($H_2O$) in order to study the liquid permeability.

In the case of our mixed materials, it was observed that the degradation of the material based on zinc and magnesium break the inter-linkages between the ceramic powders and creates a porosity on the surface and inside the material. In order to understand the influence of the porous structure on transport phenomena in the case of environmental applications for example, it is essential to find appropriate ways to quantitatively measure/describe the porous structuration. We used different characteristics (porosity, connectivity, …) that lead to obtain critical parameters able to characterize porous structures often employed in the description of many transport phenomena, such as water transmission [21].

Thus, the present study is focused on the characterization of the modified porous kaolin based on ceramic materials (DD3) and on how a liquid can diffuse into the mixed materials. We used water in the experiment in order to describe the permeability phenomenon in the different samples considering their inner structural characteristics.

## 2. Materials and methods

## 2.1 Materials

In order to prepare the studied pellets, the mullite-zircon components were made with a primary local kaolin clay-based material (DD3) obtained from the region of Guelma (Djebel Debagh) in Algeria and supplies by ETER (ceramic company in Guelma, Algeria). The crystallographic structure of kaolinite belongs to the C1 space group. Two types of hydroxyl groups are bonded to the octahedral Al cation: (1) the inner surface hydroxyls, which interact with the siloxane groups of an adjacent layer; and (2) the inner hydroxyl O-H bonds, which are parallel to the layer. The kaolinite structure is highly asymmetric, which results in strong dipolar interactions between the layers. The raw kaolin is rich in $SiO_2$ (42%), in $Al_2O_3$ (38%) and contained only small amounts of $Fe_2O_3$, MgO, CaO, and MnO. The ratio $SiO_2/Al_2O_3$ confirms that this material is close to a pure halloysite [22]. The ceramic-based material is mixed with the addition of 38% of zirconium oxide ($ZrO_2$; 99.5% purity). Zinc acetate (($CH_3COO)_2$ Zn $2H_2O$; 99.5% purity) and magnesium acetate ($Mg(CH_3COO)_2$ $4H_2O$; 99.5%) were used as additives to these components. Sodium hydroxide (NaOH; 96% purity) was used as a solvent.

The effectiveness of these substrates and their absorbability has been controlled by choosing water.

## 2.2 Preparation of powders and substrates

The DD3 + 38% $ZrO_2$ powder was treated at a temperature of 1200°C after grinding and sieving. The obtained ceramic powder was a composite of millite-zircon (called DD3Z) after this thermal transformation.

The ceramic powders were prepared with additives. The co-precipitation method was applied [23] where different percentages of zinc acetate (37 % and 50%) (DD3Z(t) + 37%ZnO, DD3Z(t) + 50%ZnO) and magnesium acetate (50 %) (DD3Z(t) + 50%Mg) were added to the previous mentioned powder. The percentages of ZnO or Mg result in the calculation of percentage of additional compounds considering the reference mass of DD3Z(t) (100 %). Different masses of Zn and Mg where used (3g, 2g, 1g, 0.5g) referred in percentage as a function of DD3Z and optimized for finally obtaining the final studied powders.

These elements were dissolved in the presence of NaOH and well mixed with distilled water (3 g in 100 ml), and then treated at 500 °C.

Cylindrical pellets were prepared based on 1 g of this powder without and with different additives using a 20 mm matrix and a hydraulic press (0.5 tons). These samples were treated at 1000 °C for two hours to improve the hardness and compacticity. The figure 1 illustrates the used procedure for the sample fabrication and shows the fabricated pellets. The pellets have a thickness of 4.5, 3.4, 5.0 and 4.0 mm and a diameter of 19.8, 17.3, 20.1 and 17.2 mm for DD3Z(t), DD3Z(t) + 37%ZnO, DD3Z(t) + 50%ZnO, DD3Z(t) + 50%Mg respectively. The sample DD3Z(t) + 37%Mg was not considered in the study because a preliminary test with a droplet of water put on the surface of the fabricated pellet showed the liquid stayed on the surface and was unable to penetrate and flow through the material with the same range of time. The study was then focused on these four materials.

## 2.3. Characterization techniques

The structural analysis of the ceramic substrates obtained from DD3+$ZrO_2$ after treatment at 1200°C with and without addition of Zn and Mg is carried out by XRD. The patterns recorded on all the samples are shown in the figure 2. A previous study [24] has shown the mineralogical and chemical characteristics, based on X-ray diffraction (XRD) of a kaolin raw known as DD3, from eastern Algeria. From this figure, one can notice that the most intense peaks are obviously those of the ceramic phases; tetragonal zircon (JCPDS 06-0266), monoclinic zirconia (JCPDS 37-1484), orthorhombic mullite (JCPDS 15-0776) and cubic cristobalite (JCPDS 01-0424) and are dominant in the patterns for the substrate DD3Z(t). The heat treatment allows the formation of the zircon by consumption of $SiO_2$ (cristobalite) which reacts with zirconia to create micro-pores. After addition of the 37% ZnO and 50% ZnO, there are minor peaks appearing at 31.83°, 34.54° and 36.32°. This latter corresponds to the (100), (002) and (101) planes of the ZnO wurtzite phase (JCPDS 03-0891), respectively. However, the big amount of Mg used in the DD3Z (t) + 50%Mg substrate is not detected. We observe a noticeable disappearance of $ZrSiO_4$ with the significant addition of 50% magnesium at the high temperature of 1000°C, this can be explained by the interaction between these two latter ZrSiO4 and Mg, which leads to the disappearance of magnesium and the absence of any phases with emergence of their glass phase.

The characterization of the sample porosity was done with a scanning electron microscope (SEM) (JSM-6301F) by analyzing the different external surface zones of the sample and by acquiring bulk informations of the samples. A surface and transverse characterization were done. For the last characterization, the pellets were cut vertically at different positions. We then acquired images of the sample surface at different locations inside the pellets giving cross-section images. The SEM pictures allow characterizing coordination numbers or the pore connectivity and average throat lengths. The average and the relative standard deviation of the pore radius were calculated.

However, before the pore-size characterization with an image processing software, the SEM images have to be pre-processed. A thresholding procedure must be done to obtain black and white binary images. For the connected pore calculation, it is necessary to find boundaries between pores (separate connected voids) using a suitable segmentation algorithm. We used an image processing software (Matlab) to transform the image in black and white and to threshold the intensity in order to evaluate the black region corresponding to a pore.

The used algorithm for determining the pore and network characteristics consists in morphological operations (closing, opening) and thresholding of the resulting SEM images. The aim is to separate the monolithic void structure of the sample into specific pores and throats connecting to each other. Initially, two connected objects should be separated by these image processing algorithms. We used the watershed segmentation [25, 26]. The basic image corresponding to this technique is the one of a drop of water falling on a topographic relief flows towards the "nearest" minimum. Watershed provides a fast and efficient method for segmentation and separation of attached clusters of objects. So, the different steps are: image binarization, noise filtering, watershed segmentation, labelling and connectivity analysis and finally building local pore networks.

The coordination number is also called pore connectivity and shows how each pore is connected to what number of adjacent pores [27 - 28]. This parameter is a key to find the relationship between porosity and permeability of porous media.

The coordination number or connectivity of a pore network is easily obtained by analyzing the abutting pores and throats and counting the number of throats connected to each pore [29]. Basically, the coordination number is defined as the number of pore throats connected to a single pore. The pore connectivity is one of the most important parameters determining the hydraulic characteristics of the porous medium. Not-interconnected porous martial has no permeability and vice versa. This parameter is a fundamental characteristic of pore networks and has a noticeable impact on hydraulic conductance of porous rocks.

The calculation of the porosity ratio is the fraction of the total ceramic volume that is taken up by the pore space. Thus, it is a single-value quantification of the amount of space available to fluid within a specific body of porous material. The porosity is simply a fraction of the total volume, and can range between 0 and 1, typically falling between 0.3 and 0.7 for soils.

The image analysis was performed with the Matlab software and therefore it was possible to use functions of the image processing toolbox to obtain the porosity characteristics.

The time required for the spread of the liquid from the top to the bottom of the sample was measured using a CCD camera (IDS µEye) with a magnification of 100X and image acquisition software (fig. 3). A movie was recorded. In order to characterize the liquid diffusion into the porous material, the movie was sequentially analyzed image per image with Matlab. A pipette (Thermo fisher) with a variable volume and a single channel was used to deliver a droplet on the top of the pellet. A custom-made sample holder was used to mount the pellets directly beneath the CCD camera. The cross-section images of the liquid penetrating into the bulk were done by cleaving the pellet with a ceramic knife.

## 3. Results and discussion
## 3.1. Characterization of the samples

The main objective of the sample analysis with the SEM was to characterize the porous material and to analyze the zones on the pellet where the pores are distributed. The degree of porosity of the ceramic pellet types was determined by calculating the pore diameters and connections. We choose to focus our calculations at the center of the pellets. A comparison with same calculations done at the edge of the samples has shown similar results for all the chosen characteristic parameters.

The figure 4 presents SEM observations of the surface and the porosity calculation. The white surfaces correspond to pores. The figure 4 shows clearly the difference between the four samples and the role played by the zinc and the magnesium to the differences in the microstructures of the ceramic material and the influence on the control of the porosity characteristics. We note that the surface and the different cross-sections of the pellets show similar characteristics. The distribution of the pore radiuses is presented in the figure 5. The pore radiuses are larger for DD3Z(t) + 50%ZnO compared to DD3Z(t) and decrease with the reduction of Zn or the addition of Mg. The average pore diameter is 71 nm in the presence of 50% of zinc while it reaches 46 nm after the addition of the same percentage of magnesium. The addition of 50% Mg reduces the number of pores and decreases the diameter. The difference between the center and the edge of the pellet is weak less than 2 %. The minimum distance between the pore is 1.1 mm ± 0.2 mm. For all the samples the calculated distance between the centroïd of the pores is similar. So, the pore radius is changed with the addition of Mg or ZnO and these compounds do not allow an increase of the porosity except for a high doping of ZnO (50%ZnO) because these two compounds were sufficiently effective to change the morphology of the sample contrary to samples without addition (DD3Z) or even in the case of a few additions of these compounds (< 50%ZnO). The results confirm the densification nature of ZnO for modifying the porous material by increasing the pore radius (x 1.3) and reducing the porosity (x 0.8). However, Mg with a low density (< 50 %) does not influence the morphology as far the experiment with a simple drop of water on the surface demonstrated it but reduces the porosity and pore radius with a high density (50%).

This result is also confirmed with the calculation of the chord length distribution. The chord length distribution (CLD) gives the probability of having a chord length between *r* and *r* + *dr* where r and *dr* represent the interface between the pore and the solid phase. For the pure sample without addition of Zn or Mg the distribution of chord lengths is broad compared to other samples. DD3Z(t) + 50%ZnO has the minimum number of chord lengths. The CLD value is similar between DD3Z(t) and DD3Z(t) + 50%ZnO but the width is reduced for DD3Z(t) + 37%ZnO and comparable in distribution with DD3Z(t) + 50%Mg. We can conclude that a weak quantity of ZnO doping or Mg do not affect the chord length.

The average grain is larger for DD3Z(t) + 50%ZnO. In this case we observe that DD3Z(t) + 50%ZnO with the maximum grain size has also the maximum pore radius. The addition of ZnO increases the grain size and the pore radius of the sample. Inversely, the average grain size and pore radius values are reduced with Mg.

The porosity ratio is 0.2, 0.18, 0.26 and 0.16 for the DD3Z(t), DD3Z(t) + 37% ZnO, DD3Z(t) + 50% ZnO and DD3Z(t) + 50% Mg samples respectively. The porosity ratio is then similar for all the four samples except for DD3Z(t)+50%ZnO where the size of the pores is the maximum. This last different value for the pore size is due to the initial state of overlapping materials of ceramic powder and zinc, where it was found that the decomposition of zinc at high temperatures works to break the inter-linkages between the ceramic powders and creates spaces on the surface of the sample and inside [30].

The pore coordination number refers to the segmented representation of the pore space and measures the number of pores with which a given pore shares throat vowels. The connectivity increases with the size of pore throats and with the increasing number of pore throats surrounding each pore. The number of pore throats that is connected with each pore is the coordination number. The throat number for all the samples is evaluated in the goal to have a knowledge of the connectivity of the pores for all the samples (fig. 6). The throats surrounding each pore have been calculated. DD3Z(t) + 50%ZnO has the best connectivity of the pores. In this material each pore is linked to 5 throats. DD3Z(t) + 50%ZnO has the minimum connections to throats. So, for this material there is a compromise between large pore radiuses and low connectivity. DD3Z(t) + 50% Mg has the smallest pore radius but the best connectivity. The average throat radius is maximum for DD3Z(t) + 50% Mg (1050 nm) and minimum for both DD3Z(t) + 50%ZnO and DD3Z(t) (702 nm). The connectivity, the pore throats and the pore diameter act on the permeability of the material. This permeability is a fundamental characteristic of pore networks and has a noticeable impact on hydraulic conductance of porous ceramics. The figure 6 represents for the four samples the topologically networks that realistically represents the pore connected by the pore throats. The pore and throat scale network modeling can predict multiphase flow properties with the wetting condition. The permeability and capillary pressure functions define the flow behavior in the four porous media. These two parameters critically depend on the geometry and topology of the pore space, the physical relationship between ceramic grains and the fluids. The permeability K is a measure of how easily fluids can flow through a given porous material and can be approximated with the equation [31, 32]:

$$K = Cr^2 \qquad (1)$$

However, empirically the permeability has been described by the following equation:

$$\log(K) = a.\varphi^b.N_{2D}^c + d \qquad (2)$$

in which a, b, c, and d are empirical constants, K is absolute permeability (md), $\varphi$ is the experimental porosity and $N_{2D}$ is the average 2-D coordination number obtained from the image analysis. Porosity and average 2D coordination numbers are efficient parameters for predicting the permeability of carbonates [33] and can also be used for predicting the permeability of the DD3Z(t) material and doped DD3Z(t) considering the influence of the two main parameters. Based on the previous characterizing parameters obtain with the 2D SEM observations we can predict qualitatively the permeability of each sample. We observe that DD3Z(t) + 50%Mg has the minimum porosity ratio but the maximum coordination number and maximum throat length. The pore radius is also the minimum among all the samples. So, this material should have a good permeability. DD3Z(t) + 50%ZnO is the opposite situation with the maximum of pore radius and porosity but an average coordination number and throat number very small. We observe also that for this material the chord length distribution was very broad, and all the different sizes of pore radius can be found in this material.

### 3.2. Liquid saturation of the pellets

In order to determine the ability of these samples to absorb liquids and to determine the time required to reach the saturation level of the filtration inside the sample we have used the experimental set-up with a CCD camera presented in the figure 3 to monitor the backside of the sample and the cross-section. This experiment of liquid transport in doped ceramic materials combines the characteristics of pore diameter, pore density and pore connectivity.

Water was used in the experiment as a simple reference liquid with a low viscosity able to test the permeability of the materials. In a first experiment, successive drops of distilled water were put on different zones along the pellet surface area to calculate the filtration time as a function of droplet locations on the pellet. The distance is considered from the edge of the pellet. The goal was to analyze the filtration parameters as a function of the homogeneity of the surface and particularly to compare the edge of the pellet with the central part.

The analysis of SEM images of the surface and the cross-section of the pellets between the edge and the center allows to compare the distribution of pores for the different samples but not the homogeneity of the material and particularly the boundaries of the pellet. In this experiment, distillated water droplets were put in different places along the surface area of the sample to measure the filtration time until the complete saturation of the porous sample as a function of the location where we put the droplet on the sample. At the end of each diffusion of droplets the diameter of the diffused liquid was measured at the location where the liquid has reached the other pellet's side. A syringe was set above the sample, and successive drops of distilled water were deposited on the surface and observed with the imaging system. Each volume of drops is 0.02 ml. Successive droplets are put on the surface until the effect of the droplet diffusion appears on the lower side of the sample. The time for observing the liquid on the back surface is measured and plotted in the figure 7 for the different samples.

The measurements have confirmed that the spread of the fluid varies as a function of the different locations on the sample and therefore it confirms that the spatial distribution of pores in the samples is nonhomogeneous and an edge effect is observed (fig.7). We observe that the slope of the curve is negative for all the samples except DD3Z meaning that the conditions of liquid flows are different between the center and the edge of the pellet with probably a porosity and a tortuosity effect different. Moreover, for DD3Z(t) the homogeneity varies differently. The pellet is less compact close to the edge than at the center for this material. We can also see that some samples have a greater capacity to store different volumes. The parameters obtained by a surface and a cross-section analyze are not fully suitable parameters to describe all the hydrodynamic behaviors and to predict the permeability of the whole bulk of pellets. Average pore sizes, average pore throat sizes and average throat lengths extracted from 2D images are not homogeneous in the bulk and a ratio of 6 could be expected. All the samples except DD3Z(t) are compact at the center and the time to reach the saturation is longer than on the edge. For DD3Z(t) material it is faster to diffuse the liquid on the edge than at the center. So, edge effects can occur in the fabrication of pellets and larger diameter of pellets are requested to avoid such effects and to guarantee a better homogeneity.

### 3.3. Time-dependent fluid flow

In a second experiment we have measured the time-dependent position of the visible wetted front at it moved through the cross-section of the pellet for the four different samples with time-lapse videos. We observed in the figure 8 that the results of the hydrodynamic flow agree with the expected results obtained with the suitable parameters measured with SEM images (pore radius, connectivity, porosity ratio, …). All the time dependent liquid position measurements can be plotted in logarithmic scale and we can observe the quadratic dependence of the penetration depth as a function of time. The obtained results can well be fitted with a Lucas-Washburn like curve [34 - 36] ($L^2$ = Bt) with B = 9.51, 1.73, 5.57, 9.74 for DD3Z(t), DD3Z(t) + 37%Zn, DD3Z(t) +50%ZnO, DD3Z(t) + 50%Mg respectively. This equation refers to a steady process, where the capillary force is compensated by gravity and viscous drag. B is a dimensionless permeability coefficient of the material. The coefficient accounts for the tortuosity of the flow path, the porosity of the material and the lengths of the pores. The flow of the liquid is generated by a negative pressure which pulls the fluid toward unwetted regions of the doped DD3 material. The liquid sucked into the porous material is accelerated by capillary force and soon thereafter the capillary force is compensated by the viscous drag so that a quasi-steady state is achieved.

The Lucas-Washburn equation can be written as:

$$L^2 = \frac{D.\gamma.\cos\theta}{4\eta} t \qquad (3)$$

where $t$ is the time for a liquid of dynamic viscosity $\eta$ and surface tension $\gamma$ to penetrate a distance $L$ into the capillary whose pore diameter is $D$ and $\theta$ the contact angle between the liquid and the sample surface. For water, $\eta = 1.10^{-3}$ kg/m.s, $\gamma = 73 \cdot 10^{-3}$ N/m. The contact angles are $\theta$ = 20.5°, 16°, 25.1°, 34.5° for DD3Z(t), DD3Z(t) + 37%ZnO, DD3Z(t) +50%ZnO, DD3Z(t) + 50%Mg respectively.

DD3Z is the material with a high penetration depth contrary to DD3Z(t) + 37%ZnO which has a very low penetration depth due to a very small porosity ratio and an average pore radius. We point out that DD3Z(t) + 50%Mg has a very fast penetration depth and reach a saturation faster which is confirmed by the logarithm plot due to low porosity ratio and a small average pore radius but with a high coordination number and a high average throat length. This measure confirms that these two parameters are very important for the permeability in porous materials. We can observe that for all the materials there is an oscillatory regime in the height around the equilibrium position, as it has been largely discussed in [37]. This effect is due to the low viscosity of the liquid. It was demonstrated that a wetting liquid put in contact with a tube indefinitely oscillates in this tube (transforming surface energy in kinetic and gravitational energies [38].

The limit in time of the linear regime and the time of the quasi-state is different for the four materials and we see that for DD3Z(t) + 50%Mg this limit occurs quickly. This limit occurs at a time $\gamma D^2/4\eta$ which can be physically understood as the time necessary for the viscous boundary layer to diffuse on a length of order fixed by the porosity parameters of the materials. The viscous drag can induce damping oscillations. The amplitude of capillary oscillations decreases with the capillary lengths as it shown for DD3Z(t) + 37%ZnO and DD3Z.The figure 8 shows that the important effect of penetration depth as a function of the time can be connected to the important role of the open porosity with the different doping. The maximum time for penetration of a water droplet is reached for DD3Z(t) + 50% Mg. DD3Z(t) + 50%ZnO and DD3Z(t) + 50%Mg seems the best material with a compromise between the penetration depth and the time for the hydrodynamic flow or permeability. Improvements in photoactivity are influenced by the template dimension, which correlates to the pore parameters of the studied material [39].

A test has been done to measure the maximum number of droplets before liquid saturation of the pellet. A number of droplets of 75 has been drop off corresponding to 1.1 ml. DD3Z(t) + 50%ZnO is the fastest with the same volume of water before saturation. This saturation effect cannot be deduced with the measured parameters obtained with the SEM images of the surface. The figure 9 confirms the differences between the doped materials. In the experiment we measured the number of drops before a complete saturation. We define the saturation as the moment where the liquid appears on the other face of the pellet. We measured the number of drops for reaching this observation. DD3Z(t) + 50%Mg is the fastest material reaching this state. The saturation occurs quickly due to the large pore coordination number and pore throat. DD3Z(t) + 50%ZnO is the longest to reach saturation because of a low pore coordination number and a small pore throat even if the pore radius is big compare to the other materials.

## Conclusions

Different types of pellets based on ceramics (mullite –zircon) and zinc /magnesium prepared by a co-precipitation method were studied. The SEM results show the structure of these pellets and the changes in composition, explained by an increase of the open porosity especially with the addition of zinc (type DD3Z(t) + 50% ZnO). The characteristics of the materials (porosity, chord length, pore throat) have been calculated. In order to confirm the influence of the materials on the porosity, the samples were studied with water. DD3Z(t) + 50%Mg is the material with the faster permeability confirmed by the high speed for liquid saturation and DD3Z(t) + 37%ZnO the slower. The method proposed allows to determine the general conditions of liquid penetration inside the ceramic sample by considering local and non-local parameters.

## Acknowledgment

This work has been supported by the Laboratory of Active Components and Materials (LACM) of Larbi Ben M'hidi University - Oum El Bouaghi, Algeria and the laboratory of MOLTECH-Anjou, University of Angers, France.

**Tables**

Table.1. parameters of the samples measured with surface and cross-section images of the pellets

| Pellets | Porosity ratio [%] | Coordination number | Average throat radius [nm] | Pore radius [nm] |
|---|---|---|---|---|
| DD3Z (t) | 20 | 4.8 | 702 | 59 |
| DD3Z(t)+37% ZnO | 18 | 4.9 | 820 | 47 |
| DD3Z(t)+50% ZnO | 26 | 4.7 | 702 | 80 |
| DD3Z(t)+50% Mg | 16 | 5.0 | 1050 | 46 |

**Figure captions**

Figure1: The step by step method for the preparation of pellets of ceramic based materials with a mixture of mullite and zircon on the top and strengthened by the addition of zinc and magnesium sorbents at different rates.

Figure 2: Analysis of XRD diagrams of the different materials with 2θ from 15° to 45°. (M) mullite, (Zr) $ZrO_2$, (ZrS), $ZrSiO_4$, (C) cristobalite and (ZO) ZnO. The mineralogical and chemical characteristics of the raw material known as DD3, from eastern Algeria could be found in [16] and [24].

Figure 3: Set-up for the characterization of the liquid diffusion into the ceramic samples.

Figure 4: SEM images of DD3Z(t) (a), DD3Z(t)+37%Zn (b), DD3Z(t)+50%Zn (c) and DD3Z(t)+50%Mg (d) with the corresponding calculated porosity network.

Figure 5: Chord length distribution for each sample with the corresponding superposition of the fitting curves. All the fitting curves are superposed for comparison.

Figure 6: Representation of the pore throat for the different samples with DD3Z(t)(a), DD3Z(t)+37%Zn (b), DD3Z(t)+50%Zn (c) and DD3Z(t)+50%Mg (d).

Figure 7: Measured time of water droplet diffusion into the sample for different positions of the drop on the pellet.

Figure 8: Permeability of the different samples with the penetration depth measurements of water as a function of time in a linear scale.

Figure 9: Time required for a complete saturation of water into the different samples measured with the number of drops until saturation.

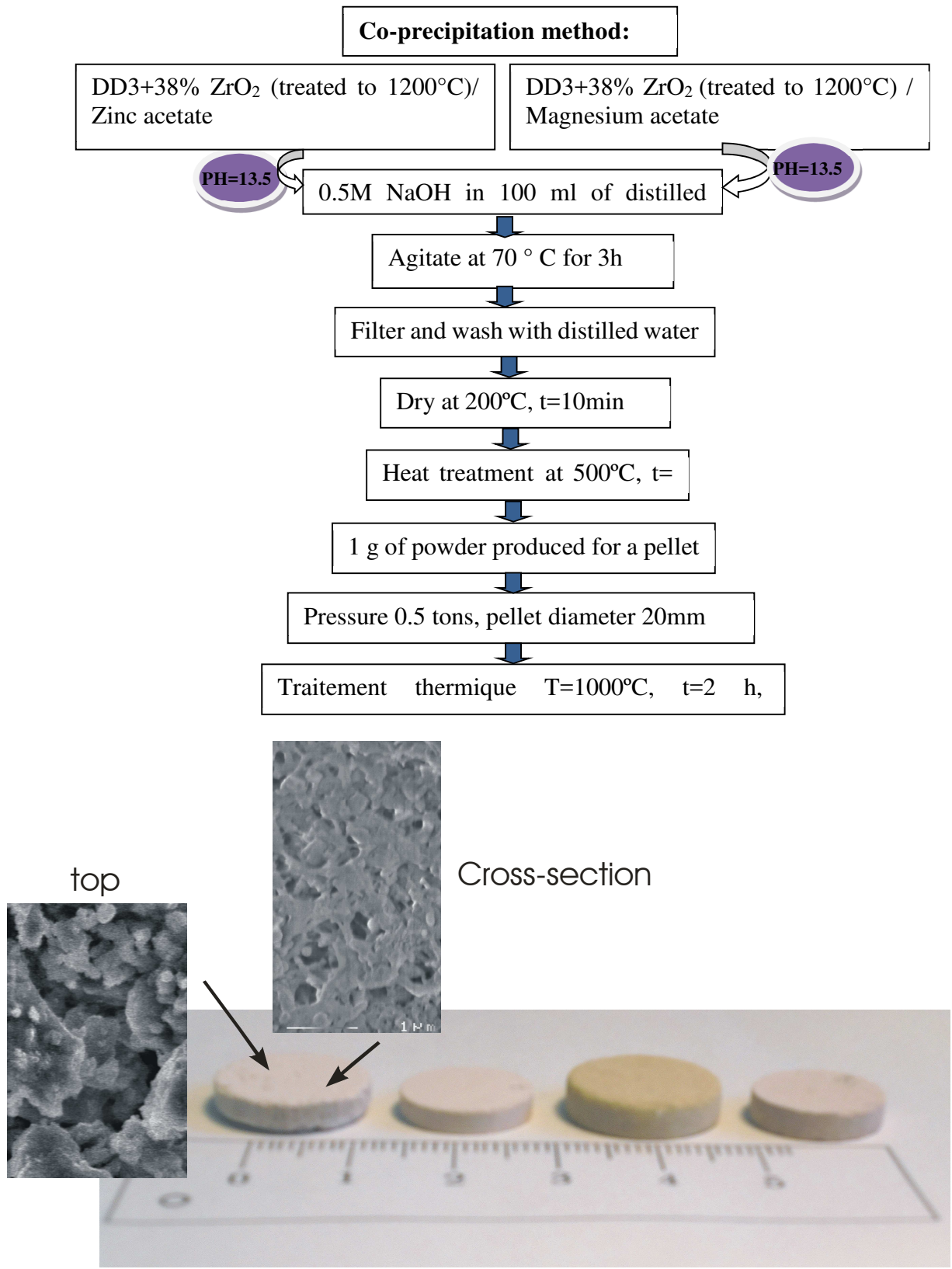

**Figure 1**

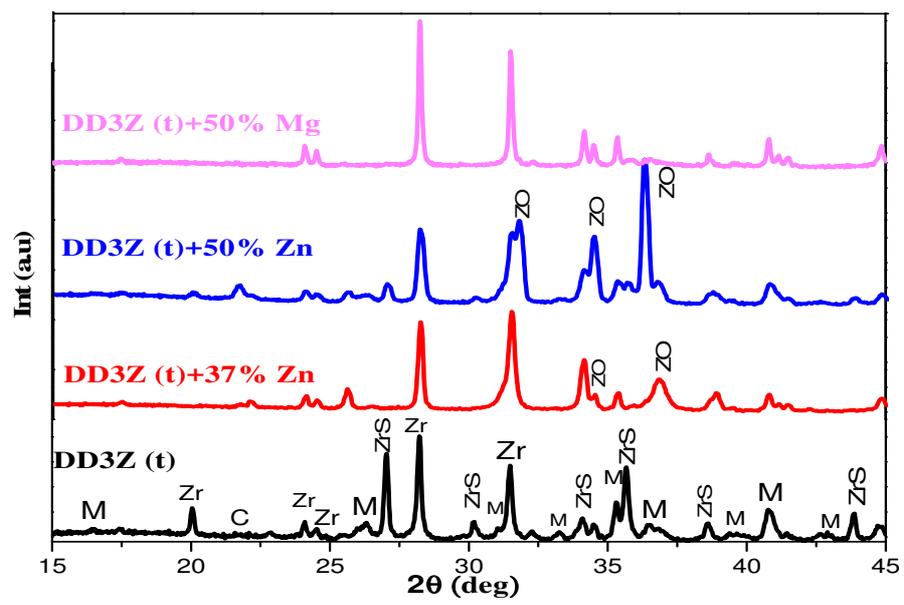

**Figure 2**

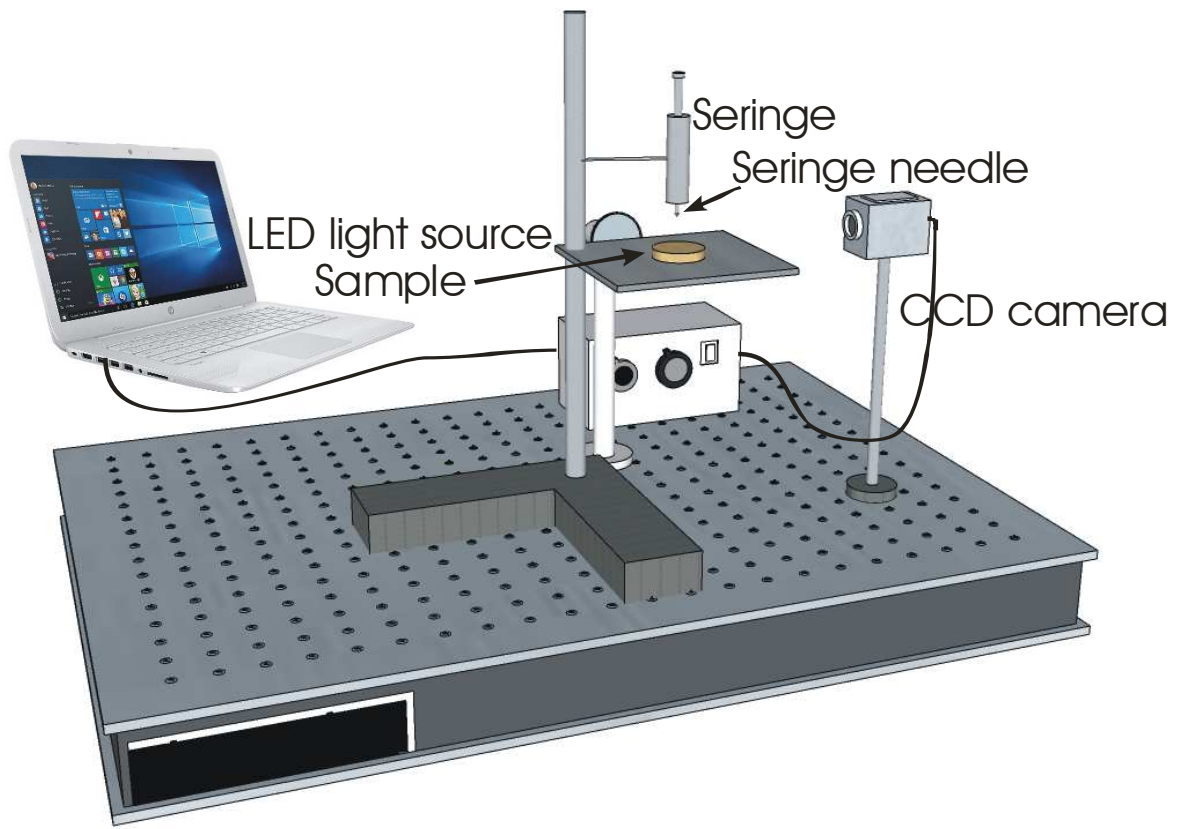

**Figure 3**

| SEM image | Porosity |
|---|---|
| 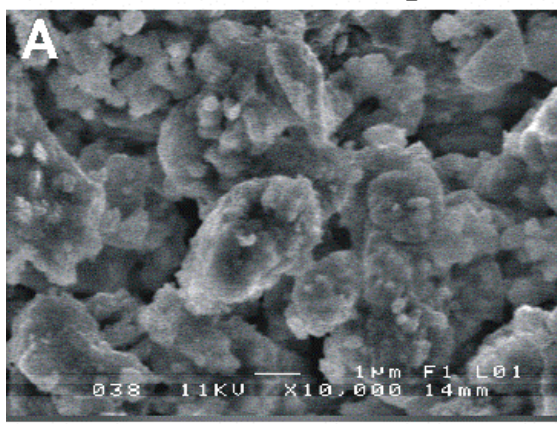 | 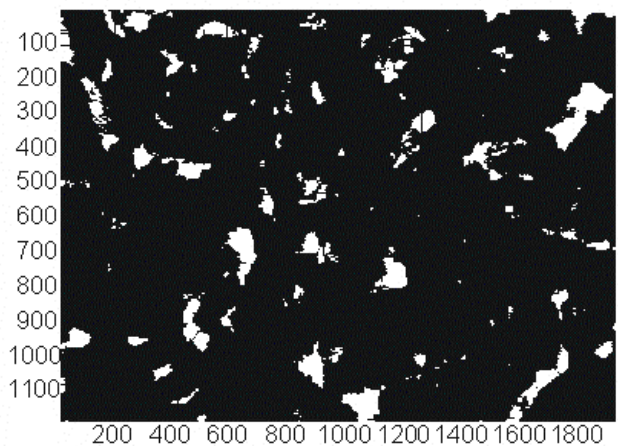 |
| 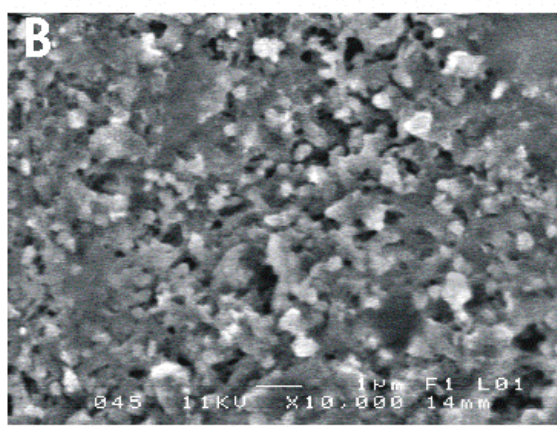 | 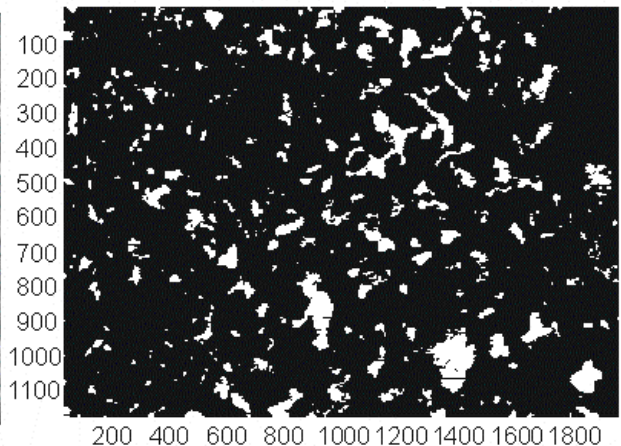 |
| 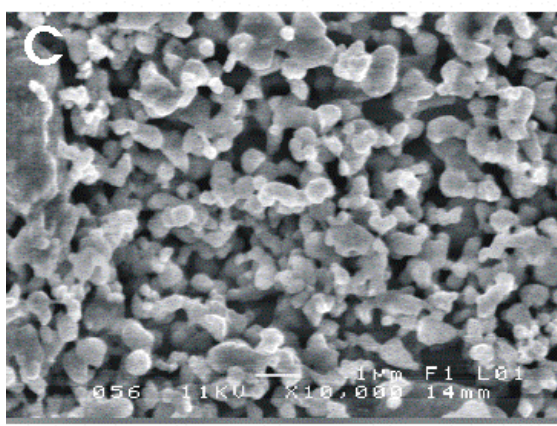 | 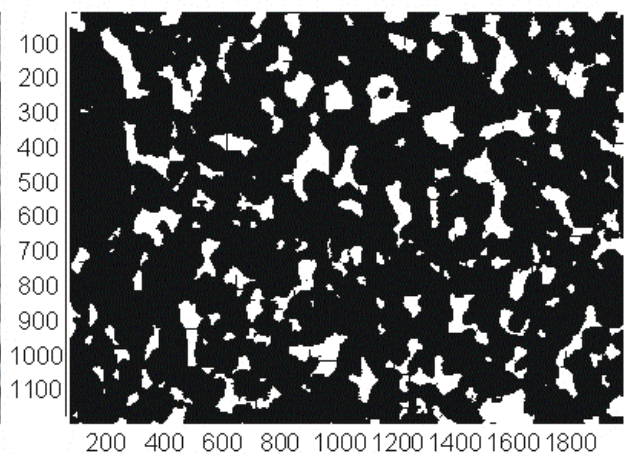 |
| 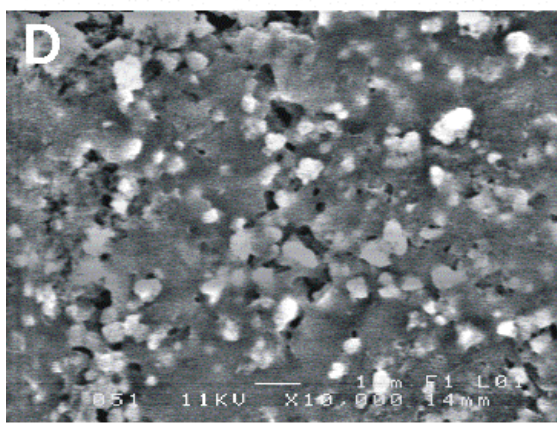 | 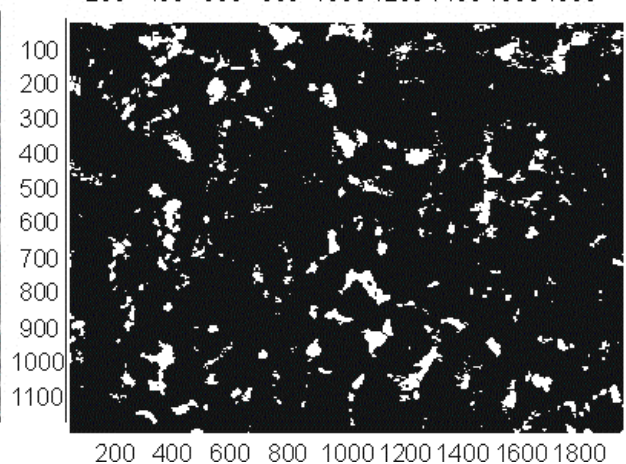 |

**Figure 4**

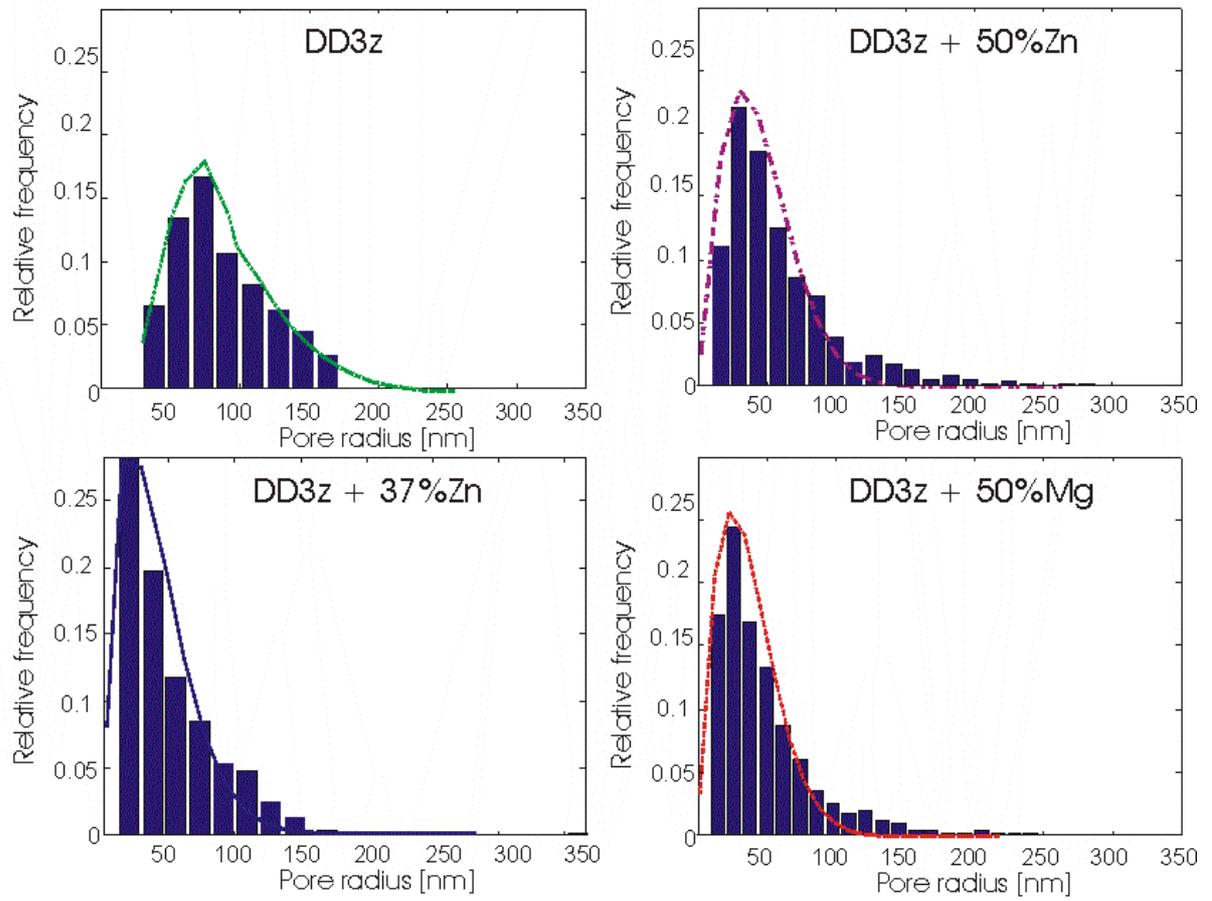

**Figure 5**

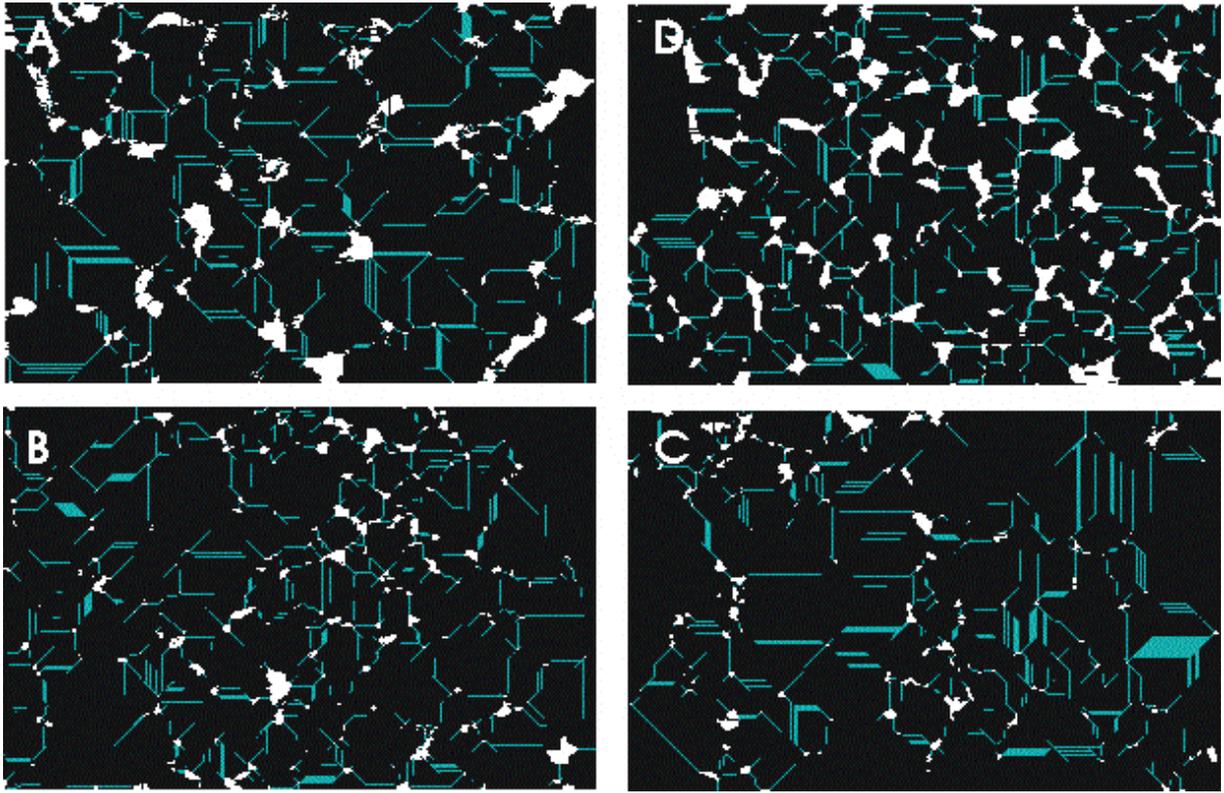

**Figure 6**

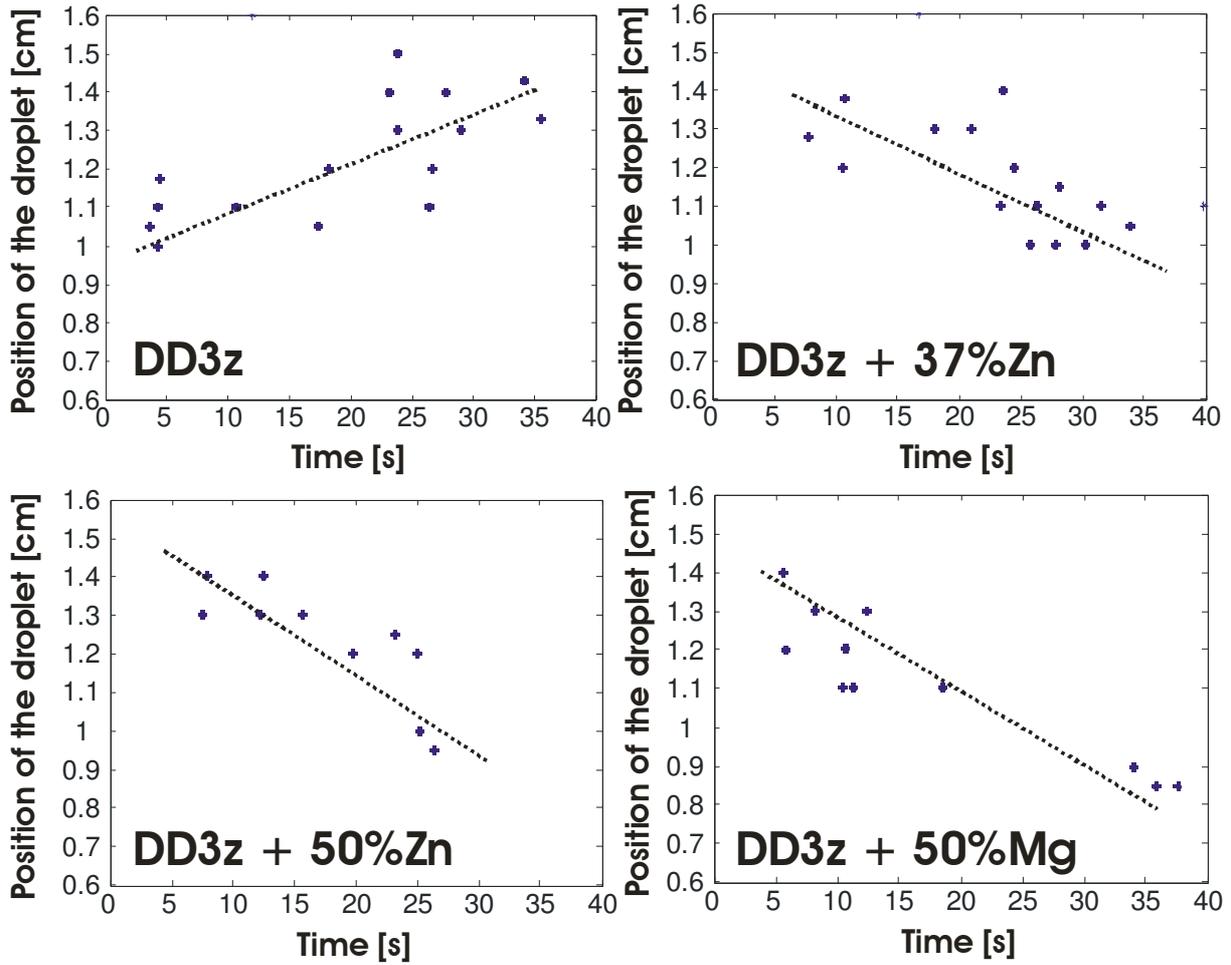

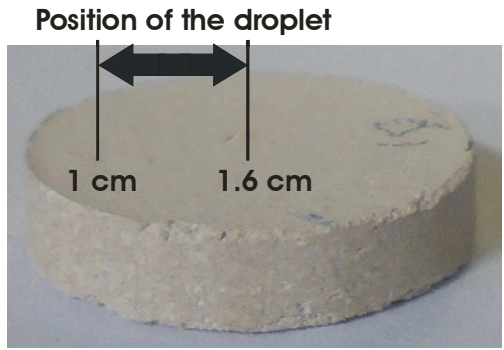

Figure 7

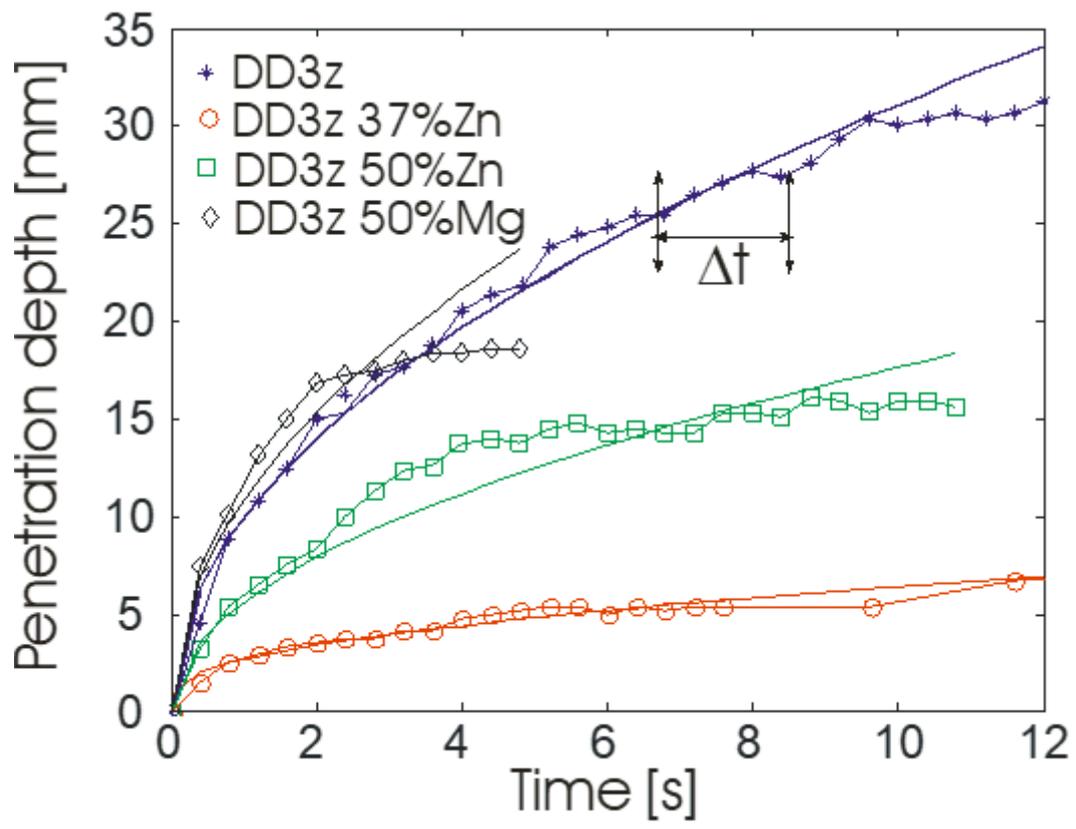

Figure 8

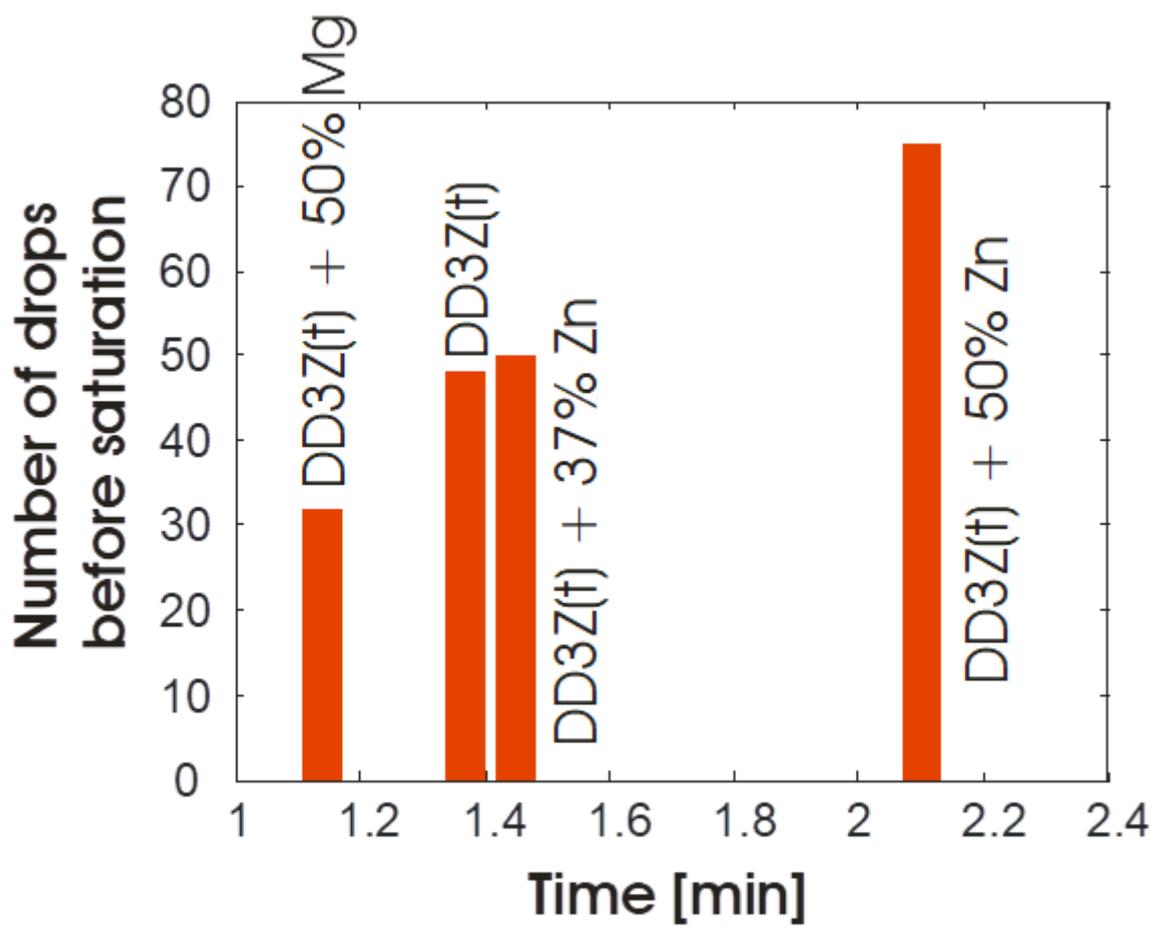

Figure 9